\documentclass[conference,hidelinks]{IEEEtran}
\IEEEoverridecommandlockouts
\usepackage{cite}
\usepackage{amsmath,amssymb,amsfonts}
\usepackage{algorithmic}
\usepackage{graphicx}
\usepackage{textcomp}
\usepackage{booktabs}
\usepackage{xcolor}
\usepackage[caption=false,font=normalsize,labelfont=sf,textfont=sf]{subfig}
\usepackage{diagbox}
\usepackage{multirow}
\usepackage[acronym]{glossaries}
\usepackage{siunitx}
\usepackage{orcidlink}
\usepackage[capitalise]{cleveref}
\usepackage{stfloats}
\usepackage{fancyhdr}

\def\bmvaOneDot{.\ }
\def\ie{\emph{i.e.}}
\def\eg{\emph{e.g}\bmvaOneDot}

\def\etal{\emph{et al}\bmvaOneDot}

\fancypagestyle{FirstPage}{
\cfoot{\footnotesize © 2024 IEEE.  Personal use of this material is permitted.  Permission from IEEE must be obtained for all other uses, in any current or future media, including reprinting/republishing this material for advertising or promotional purposes, creating new collective works, for resale or redistribution to servers or lists, or reuse of any copyrighted component of this work in other works.} 
}

\begin{document}

\title{
On-Device Domain Learning for Keyword Spotting on Low-Power Extreme Edge Embedded Systems
}

\author{
Cristian Cioflan\textsuperscript{$\dagger$}\orcidlink{0000-0003-3243-4551}, 
Lukas Cavigelli\textsuperscript{$\ddagger$}~\orcidlink{0000-0003-1767-7715}, 
Manuele Rusci\textsuperscript{$\|$}~\orcidlink{0000-0001-7458-4019}, 
Miguel de Prado\textsuperscript{$\mathparagraph$}~\orcidlink{0000-0003-4350-1617}, 
Luca Benini\textsuperscript{$\dagger$}\textsuperscript{$\mathsection$}~\orcidlink{0000-0001-8068-3806} \\
\IEEEauthorblockA{\textit{
\textsuperscript{$\dagger$} Integrated Systems Laboratory, ETH Zurich; 
\textsuperscript{$\ddagger$} Zurich Research Center, Huawei Technologies;}
} 
\IEEEauthorblockA{\textit{
\textsuperscript{$\|$} ESAT, KU Leuven;
\textsuperscript{$\mathparagraph$} VERSES AI;
\textsuperscript{$\mathsection$} DEI, University of Bologna;
}}
\IEEEauthorblockA{\small  \{cioflanc, lbenini\}@iis.ee.ethz.ch, lukas.cavigelli@huawei.com, manuele.rusci@esat.kuleuven.be, miguel.deprado@verses.ai}
}

\maketitle
\begin{abstract}
Keyword spotting accuracy degrades when neural networks are exposed to noisy environments.
On-site adaptation to previously unseen noise is crucial to recovering accuracy loss, and on-device learning is required to ensure that the adaptation process happens entirely on the edge device.
In this work, we propose a fully on-device domain adaptation system achieving up to 14\% accuracy gains over already-robust keyword spotting models.
We enable on-device learning with less than \qty{10}{\kilo \byte} of memory, using only 100 labeled utterances to recover 5\% accuracy after adapting to the complex speech noise.
We demonstrate that domain adaptation can be achieved on ultra-low-power microcontrollers with as little as \qty{806}{\milli \joule} in only \qty{14}{\second} on always-on, battery-operated devices.

\end{abstract}



\thispagestyle{FirstPage}
\newacronym{lpwan}{LPWAN}{Low-Power Wide Area Network}
\newacronym{lora}{LoRa}{Long Range}
\newacronym{lorawan}{LoRaWAN}{Long Range Wide Area Network}
\newacronym{nbiot}{NB-IoT}{Narrow Band Internet-of-Things}
\newacronym[plural=WANs, firstplural={Wide Area Networks (WANs)}]{wan}{WAN}{Wide Area Network}
\newacronym[plural=WSNs, firstplural={Wireless Sensor Networks (WSNs)}]{wsn}{WSN}{Wireless Sensor Network}
\newacronym{simd}{SIMD}{Single Instruction Multiple Data}
\newacronym{os}{OS}{Operating System}
\newacronym{ble}{BLE}{Bluetooth Low-Energy}
\newacronym{wifi}{Wi-FI}{Wireless Fidelity}
\newacronym[plural=DVS, firstplural={Dynamic Vision Sensors (DVS)}]{dvs}{DVS}{Dynamic Vision Sensor}
\newacronym{ptz}{PTZ}{Pan-Tilt Unit}

\newacronym[plural=FLLs,firstplural=Frequency Locked Loops (FLLs)]{fll}{FLL}{Frequency Locked Loop}
\newacronym{dram}{DRAM}{Dynamic Random Access Memory}
\newacronym{fpu}{FPU}{Floating Point Unit}
\newacronym{dma}{DMA}{Direct Memory Access}
\newacronym[plural=LUTs, firstplural={Lookup Tables (LUTs)}]{lut}{LUT}{Lookup Table}
\newacronym[plural=FPGAs, firstplural={Field Programmable Gate Arrays (FPGAs)}]{fpga}{FPGA}{Field Programmable Gate Array}
\newacronym{dsp}{DSP}{Digital Signal Processing}
\newacronym{mcu}{MCU}{Microcontroller Unit}
\newacronym{spi}{SPI}{Serial Peripheral Interface}
\newacronym{cpi}{CPI}{Camera Parallel Interface}
\newacronym{fifo}{FIFO}{First-In First-Out Queue}
\newacronym{uart}{UART}{Universal Asynchronous Receiver-Transmitter}
\newacronym{raw}{RAW}{Read-After-Write}
\newacronym[plural=ISAs, firstplural={Instruction Set Architectures (ISAs)}]{isa}{ISA}{Instruction Set Architecture}

\newacronym{ste}{STE}{Straight-Through-Estimator}

\newacronym[plural=PTUs, firstplural={Pan-Tilt Units}]{ptu}{PTU}{Pan-Tilt Unit}
\newacronym{mdf}{MDF}{Medium-density fibreboard}
\newacronym{cvat}{CVAT}{Computer Vision Annotation Tool}
\newacronym{coco}{COCO}{Common Objects in Context}
\newacronym{soa}{SoA}{State of the Art}
\newacronym{sf}{SF}{Sensor Fusion}

\newacronym{dl}{DL}{Deep Learning}
\newacronym{bn}{BN}{Batch Normalization}
\newacronym{FGSM}{FBK}{Fast Gradient Sign Method}
\newacronym{lr}{LR}{Learning Rate}
\newacronym{sgd}{SGD}{Stochastic Gradient Descent}
\newacronym{gd}{GD}{Gradient Descent}

\newacronym{sta}{STA}{Static Timing Analysis}

\newacronym[plural=GPIOs, firstplural={General Purpose Inupt Outputs (GPIOs)}]{gpio}{GPIO}{General Purpose Input Output}
\newacronym[plural=LDOs, firstplural={Low Dropout Regulators (LDOs)}]{ldo}{LDO}{Low Dropout Regulator}

\newacronym{inq}{INQ}{Incremental Network Quantization}

\newacronym{CV}{CV}{Computer Vision}
\newacronym{EoT}{EoT}{Expectation over Transformation}
\newacronym{RPN}{RPN}{Region Proposal Network}
\newacronym{TV}{TV}{Total Variation}
\newacronym{NPS}{NPS}{Non-Printability Score}
\newacronym{STN}{STN}{Spatial Transformer Network}
\newacronym{MTCNN}{MTCNN}{Multi-Task Convolutional Neural Network}
\newacronym{YOLO}{YOLO}{You Only Look Once}
\newacronym{SSD}{SSD}{Single Shot Detector}
\newacronym{SOTA}{SOTA}{State of the Art}
\newacronym{NMS}{NMS}{Non-Maximum Suppression}
\newacronym{ic}{IC}{Integrated Circuit}
\newacronym{rf}{RF}{Radio Frequency}
\newacronym{tcxo}{TCXO}{Temperature Controlled Crystal Oscillator}
\newacronym{jtag}{JTAG}{Joint Test Action Group industry standard}
\newacronym{swd}{SWD}{Serial Wire Debug}
\newacronym{sdio}{SDIO}{Serial Data Input Output}

\newacronym[plural=PCBs, firstplural={Printed Circuit Boards (PCB)}]{pcb}{PCB}{Printed Circuit Board}
\newacronym[plural=ASICs, firstplural={Application Specific Integrated Circuits}]{asic}{ASIC}{Application Specific Integrated Circuit}

\newacronym[plural=BNNs, firstplural={Binary Neural Networks (BNNs)}]{bnn}{BNN}{Binary Neural Network}
\newacronym[plural=NNs, firstplural={Neural Networks}]{nn}{NN}{Neural Network (NNs)}
\newacronym[plural=SCMs, firstplural={Standard Cell Memories (SCMs)}]{scm}{SCM}{Standard Cell Memory}
\newacronym{ann}{ANN}{Artificial Neural Networks}
\newacronym{ml}{ML}{Machine Learning}
\newacronym{ai}{AI}{Artificial Intelligence}
\newacronym{iot}{IoT}{Internet of Things}
\newacronym{fft}{FFT}{Fast Fourier Transform}
\newacronym[plural=OCUs, firstplural={Output Channel Compute Units (OCUs)}]{ocu}{OCU}{Output Channel Compute Unit}
\newacronym{alu}{ALU}{Arithmetic Logic Unit}
\newacronym{mac}{MAC}{Multiply-Accumulate}
\newacronym{soc}{SoC}{System-on-Chip}

\newacronym{PGD}{PGD}{Projected Gradient Descend}
\newacronym{CW}{CW}{Carlini-Wagner}
\newacronym{OD}{OD}{Object Detection}

\newacronym{rrf}{RRF}{RADAR Repetition Frequency}
\newacronym{nlp}{NLP}{Natural Language Processing}
\newacronym{qam}{QAM}{Quadrature Amplitude Modulation}
\newacronym{rri}{RRI}{RADAR Repetition Interval}
\newacronym{radar}{RADAR}{Radio Detection and Ranging}
\newacronym{loocv}{LOOCV}{Leave-one-out cross validation}

\newacronym{bsp}{BSP}{Board Support Package}
\newacronym{ttn}{TTN}{The Things Network}
\newacronym{wip}{WIP}{Work in Progress}
\newacronym{json}{JSON}{JavaScript Object Notation}
\newacronym{qat}{QAT}{Quantization-Aware Training}

\newacronym{cls}{CLS}{Classification Error}
\newacronym{loc}{LOC}{Localization Error}
\newacronym{bkgd}{BKGD}{Background Error}
\newacronym{roc}{ROC}{Receiver Operating Characteristic}
\newacronym{frr}{FRR}{False Rejection Rate}
\newacronym{eer}{EER}{Equal Error Rate}
\newacronym{snr}{SNR}{Signal-to-Noise Ratio}
\newacronym{flop}{FLOP}{Floating-Point Operation}
\newacronym{fp}{FP}{Floating-Point}
\newacronym{fps}{FPS}{Frames Per Second}

\newacronym{gsc}{GSC}{Google Speech Commands}
\newacronym{mswc}{MSWC}{Multilingual Spoken Words Corpus}
\newacronym{demand}{DEMAND}{Diverse Environments Multichannel Acoustic Noise Database}

\newacronym[plural=SNNs, firstplural={Spiking Neural Networks (SNNs)}]{snn}{SNN}{Spiking Neural Network}
\newacronym[plural=DNNs, firstplural={Deep Neural Networks (DNNs)}]{dnn}{DNN}{Deep Neural Network}
\newacronym[plural=TCNs,firstplural=Temporal Convolutional Networks]{tcn}{TCN}{Temporal Convolutional Network}
\newacronym[plural=CNNs,firstplural=Convolutional Neural Networks (CNNs)]{cnn}{CNN}{Convolutional Neural Network}
\newacronym[plural=TNNs,firstplural=Ternarized Neural Networks]{tnn}{TNN}{Ternarized Neural Network}
\newacronym{ds-cnn}{DS-CNN}{Depthwise Separable Convolutional Neural Network}
\newacronym{rnn}{RNN}{Recurrent Neural Network}
\newacronym{gcn}{GCN}{Graph Convolutional Network}
\newacronym{mhsa}{MHSA}{Multi-Head Self Attention}
\newacronym{crnn}{CRNN}{Convolutional Recurrent Neural Network}
\newacronym{clca}{CLCA}{Convolutional Linear Cross-Attention}

\newacronym{bf}{BF}{Beamforming}
\newacronym{anc}{ANC}{Active Noise Cancellation}
\newacronym{agc}{AGC}{Automatic Gain Control}
\newacronym{se}{SE}{Speech Enhancement}
\newacronym{mct}{MCT}{Multi-Condition Training}
\newacronym{mcta}{MCTA}{Multi-Condition Training \& Adaptation}
\newacronym{pcen}{PCEN}{Per-Channel Energy Normalization}
\newacronym{mfcc}{MFCC}{Mel-Frequency Cepstral Coefficient}
\newacronym{asr}{ASR}{Automated Speech Recognition}
\newacronym{kws}{KWS}{Keyword Spotting}
\newacronym{odl}{ODL}{On-Device Learning}


\newacronym{nl-kws}{NL-KWS}{Noiseless Keyword Spotting}
\newacronym{na-kws}{NA-KWS}{Noise-Aware Keyword Spotting}
\newacronym{odda}{ODDA}{On-Device Domain Adaptation}
\newacronym{hpm}{HPM}{High-Performance Mode}
\newacronym{lpm}{LPM}{Low-Power Mode}
\section{Introduction}
\label{sec:introduction}

\gls{kws} pipelines achieve high classification accuracy in clean, noiseless environments.
However, their performance severely decreases (\ie, up to 27\%~\cite{cioflan2022towards}) when the data used to train the model offline, on remote servers, differs from the in-field data, \eg due to different on-site noise conditions~\cite{cioflan2022towards}.
Works addressing this issue often remove the noise from the~\gls{dnn}'s input signal~\cite{rusci2022accelerating}, or they artificially augment utterance samples with noise during the offline training to gain noise-robust acoustic models~\cite{lopezespejo2021novel}.
Nonetheless, in such scenarios, the devices running \gls{kws} in inference mode lack the adaptation mechanisms to specialize the deployed model on noises unseen during~\gls{dnn} training, crucial for highly non-stationary noises.

The envisioned adaptation should happen fully on-board to achieve a privacy-by-design system while also minimizing the communication energy costs~\cite{chen_deep_2019}.
The migration of the \gls{ml} pipeline to low-cost, low-power edge devices---known as the TinyML paradigm---focused initially on inference~\cite{chen2018tvm,david2021tensorflow,burrello2020dory}.
Nonetheless, \gls{odl} frameworks have been recently proposed~\cite{ren2021tinyol,lin2022ondevice,nadalini2022pulptrainlib}, enabling partial or total \gls{dnn} updates on embedded platforms in the presence of on-site, real-world data distribution shifts.
Although still an open problem due to the tight constraints (\ie, memory, storage, latency, device lifetime) within embedded systems, these preliminary approaches are poised to enable efficient post-deployment adaptation at the extreme edge.

This work demonstrates fully on-device domain adaptation of keyword spotting models in ultra-low-power systems.
We achieve accuracy gains up to 14\% compared to an offline-trained, noise-robust (herein named \gls{na-kws}) model, in previously unseen noisy conditions.
We tackle more difficult scenarios than~\cite{cioflan2022towards}, by considering environments with a \gls{snr} of \qty{0}{\decibel}.
Moreover, we propose memory- and storage-reduction strategies considering resource-constrained \gls{odl} for domain adaptation, \ie, \gls{odda}.
We thus improve by 5\% the accuracy in challenging \textit{meeting} noise conditions with read-write memory requirements as low as \qty{10}{\kilo \byte}.
Finally, we are the first to demonstrate in practice the constrained \gls{odda} on an ultra-low power, multi-core GAP9 platform, enabling on-site domain adaptation in only \qty{14}{\second} at modest costs of \qty{384}{\micro \joule} per sample.

\begin{figure*}[t]
\centering
\includegraphics[width=1\textwidth]{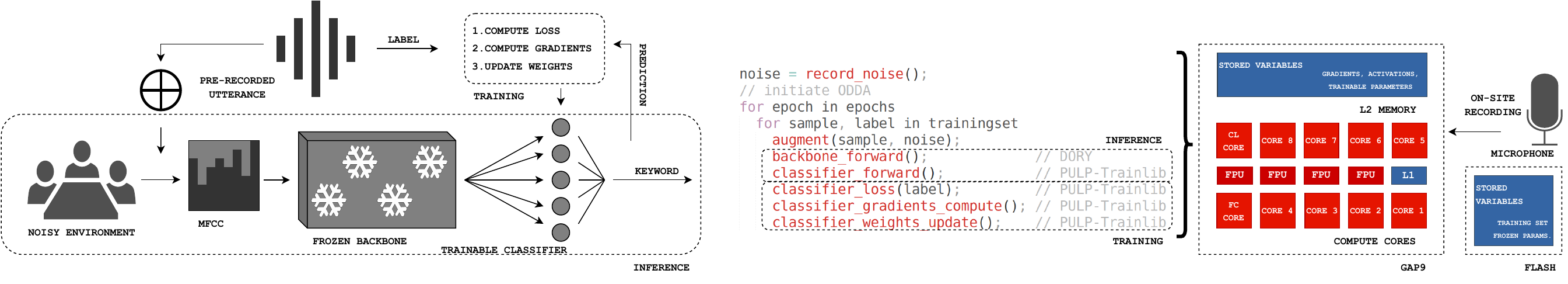}
\caption{Overview of the proposed \gls{odda} pipeline. During inference, the on-site recording is processed and its cepstral coefficients are computed, which are then fed to a frozen backbone. The classifier is used to predict the uttered keyword. During training, the on-site recording augments a prerecorded utterance. The prediction and the ground truth are jointly used to update the classifier parameters.}
\label{fig:methodology}

\vspace{-0.3cm}

\end{figure*}

\section{Related Work}
\label{sec:relatedwork}

Following the taxonomy introduced by Lopez-Espejo~\etal~\cite{lopezespejo2022deep}, environmental noise can be addressed through front-end or back-end methods. 
The former refers to strategies such as \gls{anc}~\cite{huang2018supervised, huang2019hotword} or \gls{se}~\cite{jung2020multitask,rusci2022accelerating}, which generate clean speech signal by subtracting the estimated noise component from the input signal before reaching the acoustic model.
Front-end modules increase the latency and dimension of \gls{kws} systems and do not support post-deployment model fine-tuning. 

Alternatively, back-end methods~\cite{lopezespejo2021novel,ng2022i2cr,ng2022convmixer} train acoustic models with clean utterances augmented with distortions estimated to be present in the test environment.
Such networks mitigate the front-end cost and enhance the classification performance in adverse conditions, yet they lack environmental awareness and fail to specialize on the target noise.

\Gls{odl} is therefore necessary to achieve TinyML systems specialized on real-world, on-site scenarios while respecting the privacy and security concerns of the users.
In the context of keyword spotting, \gls{odl} has been studied addressing few-shot learning, targeting smartphones and smartwatches~\cite{jagmohan2022exploring}, single-board computers~\cite{disbato2020incremental}, or \glspl{mcu}~\cite{profentzas2023minilearn, rusci2023ondevice}.
Cioflan~\etal~\cite{cioflan2022towards} proposed a domain adaptation methodology for \gls{kws} applications. 
They achieve a high energy cost of \qty{5.81}{\joule}, which would considerably reduce the lifetime of always-on, battery-operated devices, while no on-device deployment is demonstrated.

The low number of \gls{odl} \gls{kws} works is correlated with the few \gls{odl} frameworks targeting \glspl{mcu}.
Ren~\etal~\cite{ren2021tinyol} propose a TinyML training solution expanding an already deployed, frozen \gls{dnn} with a layer that can be customized through \gls{sgd}.
\cite{lin2022ondevice} introduces the \emph{Tiny Training Engine}, generating a backward computation graph enabling sparse tensor update and operator reordering on single-core ARM \glspl{mcu}.
Nadalini~\etal~\cite{nadalini2022pulptrainlib} propose PULP-TrainLib, performing matrix multiplication 36.6$\times$ faster than previous single-core solutions and thus accelerating \gls{odl}.
However, none of the previous works addressed \gls{odl} \gls{kws} for on-site noise conditions.
This paper demonstrates the first on-device domain adaptation work for \gls{kws} on ultra-low-power platforms, tackling more difficult scenarios and tighter resource constraints than previous works.

\section{Background}
\label{sec:background}

We consider the \gls{odda} methodology proposed by Cioflan~\etal~\cite{cioflan2022towards}, refining a \gls{kws} network on on-site, previously unseen background noise.
A \gls{na-kws} model, which was trained on the server using a full dataset of keyword clean utterances, augmented with multiple noise types, is deployed on the target platform.
During inference, the \gls{kws} system processes incoming utterances recorded on-site.
The signal is preprocessed and fed to the \gls{dnn}, which predicts the uttered keyword, as shown in \cref{fig:methodology} (left).

The \gls{odda} process initiates once a new noise is present in the environment, noise that can potentially harm the performance of the \gls{kws} model.
First, the noise signal is recorded and used to augment clean, labeled utterances.
The utterances belong to the \gls{na-kws} training set and are stored in a non-volatile memory to relieve the user from the tedious process of manually recording (and labeling) data.
The resulting signal is fed to the model to compute the prediction and, together with the input label, we determine the cross-entropy loss.
We calculate the gradient of the loss with respect to the weights using the backpropagation algorithm and, finally, we update the model weights using~\gls{gd}.
The steps are repeated for each prerecorded training utterance stored on the board, for multiple adaptation epochs.
\section{Resource-Constrained ODDA}
\label{sec:methodology}

We refine the \gls{odda} methodology to make it suitable for TinyML settings.
Considering always-on, battery-powered target devices, we address energy consumption, which is proportional to the model's adaptation latency (t\textsubscript{ODDA}), the number of training samples, and the number of epochs, according to the relation:

\vspace{-0.4cm}

\begin{equation}
\label{eq:oddatime}
    t_\mathrm{ODDA} \propto (t_\mathrm{infer} + t_\mathrm{backprop}) * \frac{S_\mathrm{dataset}}{S_\mathrm{batch}} * N_\mathrm{epochs},
\end{equation}

where S\textsubscript{dataset} represents the on-device dataset size and S\textsubscript{batch}, the batch size. 
We denote as t\textsubscript{infer} the batch inference latency and as t\textsubscript{backprop} the latency in updating the model's parameters. 
The adaptation latency of \gls{odda} depends on the number of noise adaptation rounds (\ie, $N_\mathrm{epochs}$), as well as on the target model; a lightweight model, with few \glspl{flop}\footnote{One \gls{flop} represents one floating-point addition or multiplication.} completes the parameter update faster than a complex network. 

Another constraint associated with extreme edge devices is the available read-write, on-chip memory.
\Gls{odda} relies on \gls{gd} learning, which requires the model inputs, parameters, activations, and gradients, associated with a particular layer, to be available simultaneously in the memory.
Although the latency scales inversely with the batch size, as shown in ~\cref{eq:oddatime}, processing more samples in parallel is limited by the size of the memory containing the backpropagation-associated variables.
Lastly, we also address the storage constraints, \ie, the read-only, off-chip memory used to store the model frozen parameters, as well as the prerecorded utterances.

We propose two optimization strategies to reduce the impact of \gls{odda} on the limited available resources. 
We first enable flexible adaptation depths, partially freezing trainable parameters of the target model.
This reduces the number of parameters, activations, and gradients needed for adaptation, thus minimizing the on-chip memory requirements, while simultaneously reducing the computational requirements associated with gradient computation and weight update.
Second, we address the available storage space by performing \gls{odda} on a fraction of the keyword samples available during training time. 
The storage constraints also impact the adaptation time, as performing the same number of adaptation rounds on fewer samples (\ie, reducing S\textsubscript{dataset} in \cref{eq:oddatime}) results in fewer \glspl{flop} and, therefore, in a faster learning process. 

\section{On-Device Implementation}
\label{sec:ondeviceimplementation}

We implement the \gls{odda} methodology on the GAP9 \gls{mcu}, depicted in~\cref{fig:methodology} (right), featuring ten general-purpose RISC-V cores organized in a fabric controller and a nine-core cluster supporting 8-bit integer (\ie, \texttt{int8}) \gls{mac} instructions.
The cluster also shares four \glspl{fpu}, for half (\ie, \texttt{fp16}) and single (\ie, \texttt{fp32}) precision operations, and a hardware convolution accelerator, with energy consumption levels as low as \qty{5}{\micro \joule} for \glspl{dnn}~\cite{tinybenchmark}.
GAP9 features a hierarchical memory architecture, with an L1 TCDM of \qty{128}{\kilo \byte}, an L2 SRAM of \qty{1.5}{\mega \byte}, and both on- and off-chip L3 Flash and RAM memories.

We use \gls{ds-cnn} as our acoustic model~\cite{zhang2017hello, lu2019depthwise, sorensen2020depthwise}, shown to obtain a competitive keyword spotting system that can function in low- and ultra-low-power platforms. 
The frozen section of the \gls{ds-cnn} model is deployed on-device using the DORY code-generator~\cite{burrello2020dory}, after quantizing its corresponding parameters to \texttt{int8}. 
We moreover store in the off-chip L3 memory prerecorded, labeled utterances used for training \gls{na-kws}.
We use PULP-TrainLib~\cite{nadalini2022pulptrainlib} to generate inference and training C code for our target, learnable layer(s), computed in \texttt{fp32} to preserve the numerical accuracy needed during training.

The \gls{odl} algorithm, shown in~\cref{fig:methodology} (center), initiates in the presence of a new noise. 
Environmental noise is recorded using a microphone, and it is used to augment the prerecorded training samples.
\texttt{fp16} \glspl{mfcc} are computed and fed to the frozen backbone.
Its dequantized output is employed, together with the utterance label, by the \texttt{fp32} classifier to compute the cross-entropy loss.
The loss gradients with respect to the weights are then calculated, followed by updating the model's learnable parameters.
The \gls{odda} algorithm therefore requires minimal effort from the end-user, while simultaneously yielding specialized \gls{kws} systems ready to be used in noisy environments.

\section{Results}
\label{sec:results}

\begin{table*}
\caption{The effect on the accuracy [\%] of increasing the task complexity by increasing the number of classes in conjunction with an increase in model capacity.}
\label{tab:dataimpact}
\resizebox{\linewidth}{!}{%
\begin{tabular}{c c | c c c c c c c c c c c c c c c}  
 \toprule
    \multicolumn{2}{r|}{Noise Type $\rightarrow$} & \multicolumn{3}{c}{--------- Cafeteria ---------} & \multicolumn{3}{c}{--------- Meeting ---------} & \multicolumn{3}{c}{-------- Restaurant --------} & \multicolumn{3}{c}{--------- Washing ---------}& \multicolumn{3}{c}{--------- Metro ---------} \\
    Model &  Mode & 6 cls. & 12 cls. & 35 cls. & 6 cls. & 12 cls. & 35 cls. & 6 cls. & 12 cls. & 35 cls. & 6 cls. & 12 cls. & 35 cls. & 6 cls. & 12 cls. & 35 cls. \\
     \midrule
    \multirow{2}{*}{S} & NA-KWS & 88.07 & 78.04 & 55.60 & 79.35 & 65.69 & 49.19 & 83.30 & 73.52 & 53.68 & 94.98 & 89.30 & 64.94 & 92.84 & 83.82 & 63.89\\
     & ODDA & 90.07 & 80.59 & 57.31 & 88.41 & 78.53 & 57.19 & 85.59 & 76.22 & 55.97 & 96.24  & 90.45 & 65.47 & 93.10 & 85.50 & 64.95 \\
     \midrule
    \multirow{2}{*}{M} & NA-KWS & 90.29 & 82.60 & 59.60 & 83.18 & 69.96 & 51.06 & 85.61 & 77.83 & 57.03 & 96.63 & 91.64 & 67.58 & 94.30 & 87.10 & 65.82\\
    & ODDA & 91.78 & 84.25 & 60.69 & 91.41 & 84.23 & 57.87 & 88.76 & 80.12 & 59.01 & 97.10 & 92.86 & 68.15 & 94.15 & 89.51 & 66.66 \\ 
     \midrule
    \multirow{2}{*}{L} & NA-KWS & 91.31 & 83.91 & 62.16 & 85.13 & 70.76 & 55.27 & 87.39 & 79.53 & 59.31 & 96.82 & 92.92 & 71.46 & 94.75 & 87.73 & 68.40 \\
    & ODDA & 92.80 & 85.54 & 63.53 & 92.89 & 84.74 & 63.49 & 88.88 & 81.55 & 61.19 & 97.22 & 93.39 & 71.97 & 95.54 & 90.06 & 69.10 \\ 
     \bottomrule
\end{tabular}
}

\vspace{-0.3cm}

\end{table*}

We consider the~\gls{gsc}~\cite{warden2018gsc} dataset as our keyword spotting task, using the data split proposed by the authors.
To model the noisy environment, we consider additive noise sourced from real-world recordings of \gls{demand}~\cite{thiemann2013demand} dataset. 
Following \cref{sec:methodology}, for~\gls{na-kws}, we use, with equal probability, 17 out of 18 noises to augment the samples, together with the absence of any noise, \ie, the  \emph{clean} environment.
The omitted noise, herein named \textit{target}, is present only after deployment, as opposed to the \textit{source} noises used for offline training. 
Our training and evaluation experiments are performed using an~\gls{snr} of \SI{0}{\decibel}, with the \gls{ds-cnn} models described in \cref{tab:gap9_measurements}.

\subsection{Specializing on noise}
\label{ssc:specializing}

In~\cref{tab:dataimpact}, we analyze the \gls{odda} impact given the task complexity and the model capacity, demonstrating that \gls{odda} enables specialization on any previously unseen noise.
The impact of~\gls{odda} varies across target environments, with improvements of 1.15\% on noises with stationary properties  (\ie, \textit{washing}), as the generally-robust \gls{na-kws} network is already capable of determining the speech features in the presence of these noises.
On the other hand, domain adaptation is imperative for non-stationary noises, such as \textit{meeting}; \gls{odda} \gls{ds-cnn} S outperforms by 12\% its \gls{na-kws} counterpart in these conditions.
Improvements as high as 15\% are measured for larger \glspl{ds-cnn}, leading to top-1 accuracies up to 93\% in \textit{washing} conditions.
Notably, in the presence of unseen speech noise, domain adaptation is crucial, as a refined \gls{ds-cnn} S outperforms the 17$\times$ larger \gls{na-kws} \gls{ds-cnn} L by 8\%.

When the task complexity increases to a 35-class problem, \gls{odda} brings accuracy improvements between 1\% (\ie, \textit{washing}) and 8\% (\ie, \textit{meeting}) over \gls{na-kws} models, noise exposure enabling networks to identify distinctive speech properties.
We further define a 6-word task (\ie, "yes", "down", "left", "right", "off", "stop") following~\cite{mazumder2021mswc}, eliminating short, phonetically similar words (\eg "no" and "go").
The \gls{odda} impact reduces with the task complexity, its benefits peaking at 11\% as it is easier to obtain feature separability even in noisy conditions. 
The \gls{odda} gains lower with the increase in model capacity, down to 7\% for \gls{ds-cnn} L, as larger models are already capable of encoding richer class representations.
This comes with unprecedented top-1 accuracies, with the largest models scoring between 88\% and 97\% in noisy conditions.

\begin{figure*}[t]
\centering
{\includegraphics[width=0.49\textwidth]{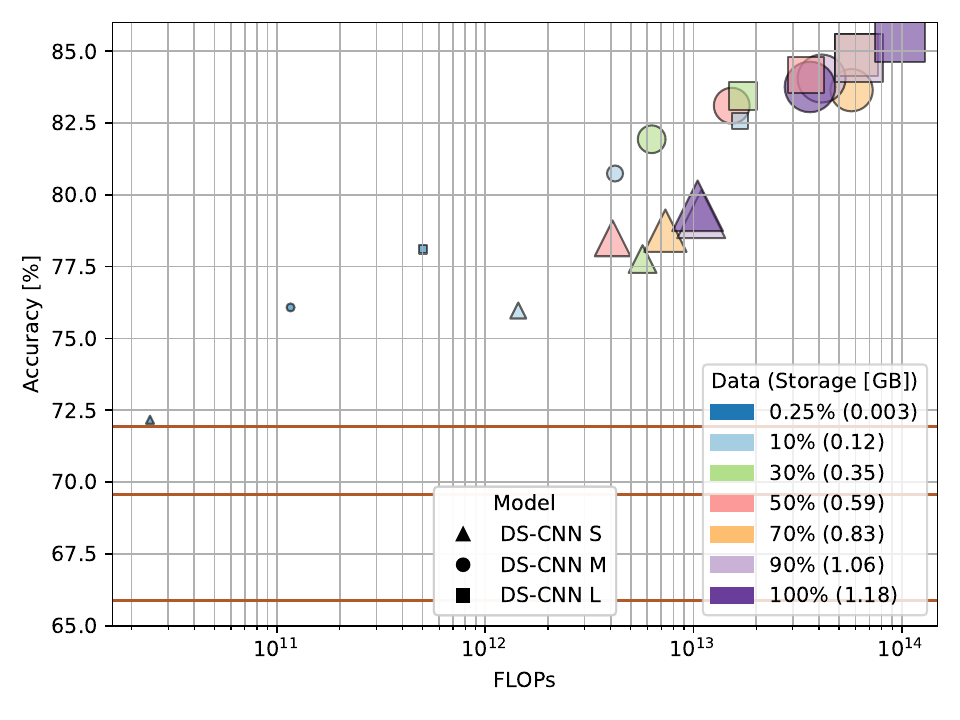}\label{fig:dataxmodelsize}}
\hfill
{\includegraphics[width=0.49\textwidth]{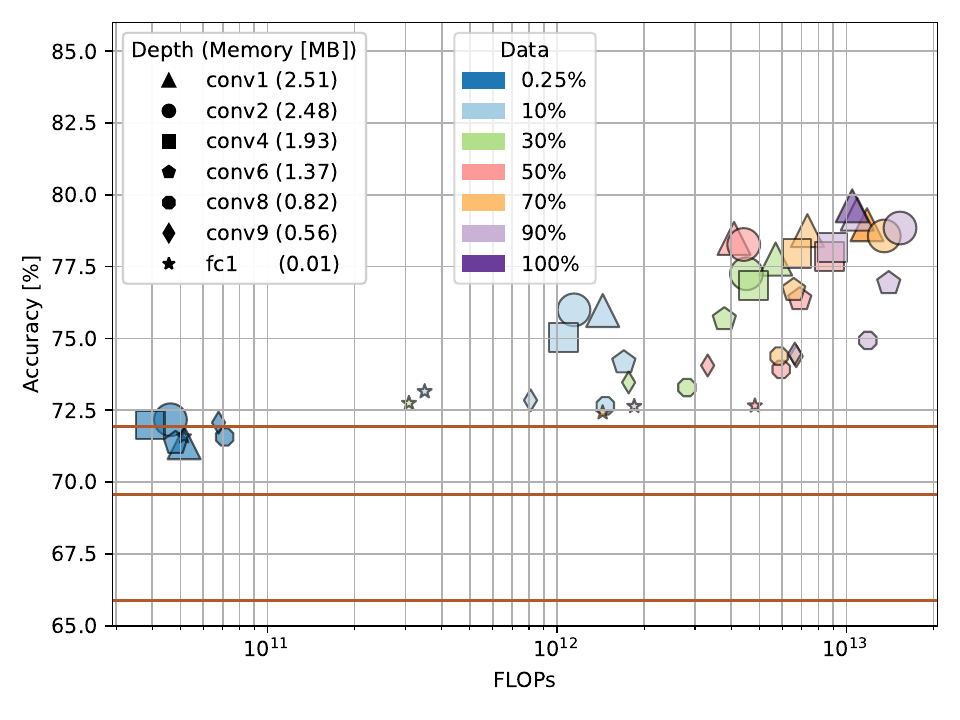}\label{fig:dataxlayers}}

\vspace{-0.4cm}

\caption{Mitigating resource constraints during ODDA considering \textit{speech} noise. (left) represents the storage cost given the \gls{ds-cnn} model (\ie, S, M, L) and the amount of randomly selected data used, while (right) shows the memory cost given the number of updated layers of \gls{ds-cnn} S, starting with the last linear (\textit{fc1}) layer, for a batch size of two. The horizontal brown lines represent, starting with the lowest one, the baseline accuracy obtained with \gls{na-kws} \gls{ds-cnn} S, M, and L.}
\label{fig:datax}

\vspace{-0.7cm}

\end{figure*}

\subsection{Resource-constrained ODDA}
\label{ssc:resourceconstrained}

In \cref{fig:datax} (left) we represent the storage-latency trade-offs for the optimization strategies proposed in \cref{sec:methodology}.
We focus on speech noise scenarios (\ie, \textit{meeting}), where \gls{odda} is essential. 
We consider the number of~\gls{gsc}12 samples stored, requiring \qty{32}{\kilo \byte} of storage space per sample.
With only \qty{3}{\mega \byte} of data, the adapted \gls{ds-cnn} S (\ie, \SI{26}{\kilo \byte}) model surpasses its \gls{na-kws} counterpart by 6\%, achieving comparable results with the largest \gls{na-kws} \gls{ds-cnn} L.
When 10\% of the data becomes available (\ie, \SI{120}{\mega \byte} -- 310 samples per class), \gls{odda} \gls{ds-cnn} S outperforms the largest noise-robust network by 10\%, whereas employing the complete training set (\ie, \SI{1.1}{\giga \byte} for 37000 samples) generates accuracy gains up to 14\%, demonstrating the positive impact of on-site noise-augmented, prerecorded utterances in \gls{odda}.
The model size and the amount of data used also impacts the total number of \glspl{flop} performed during \gls{odda} (\ie, from \qty{1.43}{\tera \glspl{flop}} with 10\% of the data with \gls{ds-cnn} S to \qty{10.46}{\tera \glspl{flop}} when all the samples are used, and up to \qty{98.10}{\tera \glspl{flop}} in the same scenario for \gls{ds-cnn} L).
Although larger models achieve higher accuracies, their complexity increases the inference latency and slows down adaptation.

To address the memory requirements, we only update the last \textit{k} layers of the model, depicted in \cref{fig:datax} (right).
Notably, we outperform the \gls{na-kws} \gls{ds-cnn} S baseline by 5.5\% by refining only its last linear layer (\ie, \textit{fc1}), achieving comparable accuracies with \gls{na-kws} \gls{ds-cnn} L. 
This requires storing 780 trainable parameters, the input and output activations, and the weight gradients, \ie, no more than \qty{10}{\kilo \byte} of on-chip memory.
The accuracy increases with the adaptation depth and the number of utterances employed.
Using 10\% of available data, refining the entire network outperforms by 1.2\% a model of which only the classifier was adapted, whereas the difference increases to 6\% for 100\% of training samples.

\begin{table}
\caption{On-Device Domain Adaptation cost per sample in Low-Power Mode (LPM; \qty{240}{\mega \hertz}, \qty{650}{\milli \volt}) and High-Performance Mode (HPM; \qty{370}{\mega \hertz}, \qty{800}{\milli \volt}), for three \gls{ds-cnn} models. Memory refers to the read-write memory requirements.}
\label{tab:gap9_measurements}
\resizebox{\linewidth}{!}{
\begin{tabular}{c | c c c c | c c c }  
 \toprule
   DS-CNN  & Compute & Params. & Mem. & Eff.[FLOP/ &  \multirow{2}{*}{Mode} & Compute  & Energy \\
   Model & [MFLOPs] & [kB] & [kB] & cycle] & & time [ms]   & [\textmu J]  \\
   \midrule
   \multirow{2}{*}{S} & \multirow{2}{*}{2.95} &  \multirow{2}{*}{23.7} &  \multirow{2}{*}{9.5} &  \multirow{2}{*}{4.94} & LPM & 10.89 & 424 \\
                      & & & & & HPM & 6.74 & 384 \\
   \multirow{2}{*}{M} & \multirow{2}{*}{17.2} &  \multirow{2}{*}{138.1} &  \multirow{2}{*}{25.5} & \multirow{2}{*}{9.18}  & LPM & 24.16 & 988 \\
                      & & & & & HPM & 16.34 & 974 \\
   \multirow{2}{*}{L} & \multirow{2}{*}{51.1} &  \multirow{2}{*}{416.7} &  \multirow{2}{*}{40.9} & \multirow{2}{*}{11}  & LPM & 55.04 & 2313 \\
                      & & & & & HPM & 32.95 & 2028 \\
    \bottomrule 
    
\end{tabular}
}

\vspace{-0.6cm}

\end{table}
\subsection{GAP9 implementation}
\label{ssc:gap9}
Motivated by the memory- and storage-constrained \gls{odda} validated in \cref{ssc:resourceconstrained}, we implement the lightweight domain adaptation on GAP9 \gls{mcu}.
We consider 100 labeled utterances for adapting our \gls{kws} system, as described in~\cref{sec:ondeviceimplementation}.
In \cref{tab:gap9_measurements}, we depict the \gls{odda} cost per sample on GAP9 \gls{mcu} for \gls{lpm} and \gls{hpm} scenarios for the complete processing pipeline.
In \gls{lpm} we measure an average power consumption of \qty{36}{\milli \watt}, saving energy costs during inference, triggered every \qty{50}{\milli \second} in always-on, battery-operated devices, while \gls{hpm} enables energy-efficient training under \qty{55}{\milli \watt}.

We refine the classifier layer of \gls{ds-cnn} S using 100 samples (\ie, ten samples per target class) for 21 epochs, increasing the model accuracy with 6\% in only \qty{14}{\second}.
In \gls{hpm}, only \qty{806}{\milli \joule} are sufficient for effective \gls{odda}.
As larger \gls{kws} are used, memory requirements concerning classifier update increase from \qty{10}{\kilo \byte} to \qty{40}{\kilo \byte}, whereas the average energy consumption increases 5.3$\times$; up to \qty{1.9}{\min} are needed to update the largest \gls{ds-cnn} in \gls{hpm}.

\vspace{-0.4cm}
\section{Conclusion}
\label{sec:conclusion}

This work proposes a domain adaptation strategy addressing resource-constrained 
\gls{kws} systems.
We achieve accuracy gains up to 14\% for complex speech noises in 12-word problems.
For such speech background noise, only 100 prerecorded, labeled samples are sufficient to reach on-device adaptation by refining the \gls{dnn}'s classifier, with only \qty{3}{\mega \byte} of read-only memory and \qty{10}{\kilo \byte} of read-write memory.
Notably, we show that a small \gls{ds-cnn} refined through \gls{odda} outperforms 17$\times$ larger networks by up to 8\%.
We demonstrate our resource-constrained \gls{odda} on GAP9 \gls{mcu}, achieving a 6\% accuracy increase in noisy environments after only \qty{14}{\second} while requiring as little as \qty{806}{\milli \joule}.

\vspace{-0.4cm}

\section{Acknowledgements}
\vspace{-0.1cm}
This work was partly supported by the Swiss National Science Foundation under grant No 207913: TinyTrainer: On-chip Training for TinyML devices.

\pagebreak

\bibliographystyle{IEEEtran}

\bibliography{main}

\end{document}